\documentclass[acmsmall,screen]{acmart}

\usepackage[linesnumbered,lined,ruled,commentsnumbered]{algorithm2e}
\usepackage{threeparttable}
\usepackage{tcolorbox}
\usepackage{colortbl}
\usepackage{booktabs} 
\usepackage{graphicx}
\usepackage{paralist}
\usepackage{multirow}
\usepackage{multicol}
\usepackage{amsmath} 
\usepackage{cleveref}
\usepackage{xcolor}
\usepackage[dvipsnames]{xcolor}
\usepackage{xspace}
\usepackage{url}
\usepackage{array}
\usepackage[normalem]{ulem}
\usepackage{pifont}
\usepackage{enumitem}
\usepackage{minted}
\useunder{\uline}{\ul}{}

\definecolor{gray1}{rgb}{0.61, 0.75, 0.91}
\definecolor{gray1}{rgb}{0.77, 0.85, 0.95}
\definecolor{gray1}{rgb}{0.88, 0.92, 0.97}
\definecolor{gray4}{rgb}{0.86, 0.91, 0.92}
\definecolor{blue1}{rgb}{0.423, 0.557, 0.749}

\newcommand{\ie}{\textit{i}.\textit{e}.,\xspace}
\newcommand{\eg}{\textit{e}.\textit{g}.,\xspace}
\newcommand{\splvd}{SPLVD\xspace}
\newcommand{\Spl}{Self-paced learning\xspace}
\newcommand{\spl}{self-paced learning\xspace}

\setminted{
linenos, 
breaklines, 
fontsize=\scriptsize, 
numbersep=5pt, baselinestretch=1.0,
}

\AtBeginDocument{%
  }

\begin{document}

\title{Leveraging Self-Paced Learning for Software Vulnerability Detection}

\author{Zeru Cheng}
\orcid{0009-0001-5094-4427}
\email{zeru\_cheng@smail.nju.edu.cn}
\authornote{Both authors contributed equally.}
\affiliation{%
  \institution{Nanjing University}  
  \city{Nanjing}
  \country{China}
}

\author{Yanjing Yang}
\orcid{0009-0006-8789-4589}
\email{yj\_yang@smail.nju.edu.cn}
\authornotemark[1]
\affiliation{%
  \institution{Nanjing University}  
  \city{Nanjing}
  \country{China}
}

\author{He Zhang}
\orcid{0000-0002-9159-5331}
\email{hezhang@nju.edu.cn}
\affiliation{%
  \institution{Nanjing University}  
  \city{Nanjing}
  \country{China}}

\author{Lanxin Yang}
\orcid{0000-0002-0406-2263}
\email{lxyang@nju.edu.cn}
\authornote{Corresponding Author: Lanxin Yang.}
\affiliation{%
  \institution{Nanjing University}  
  \city{Nanjing}
  \country{China}}

\author{Jinghao Hu}
\orcid{0009-0001-4869-9098}
\email{jinghao\_hu@smail.nju.edu.cn}
\affiliation{%
  \institution{Nanjing University}  
  \city{Nanjing}
  \country{China}
}

\author{Jinwei Xu}
\orcid{0009-0004-5157-1118}
\email{jinwei\_xu@smail.nju.edu.cn}
\affiliation{%
  \institution{Nanjing University}  
  \city{Nanjing}
  \country{China}}

\author{Bohan Liu}
\orcid{0000-0002-0146-5411}
\email{bohanliu@nju.edu.cn}
\affiliation{%
  \institution{Nanjing University}  
  \city{Nanjing}
  \country{China}}

\author{Haifeng Shen}
\orcid{0000-0002-8221-981X}
\email{haifeng.shen@scu.edu.au}
\affiliation{%
  \institution{Southern Cross University}  
  \city{Gold Coast}
  \country{Australia}}

\begin{abstract}
Software vulnerabilities are major risks to software systems. 
Recently, researchers have proposed many deep learning approaches to detect software vulnerabilities. 
However, their accuracy is limited in practice. 
One of the main causes is low-quality training data (\ie source code). 
To this end, we propose a new approach: \textbf{\splvd} (\underline{\textbf{S}}elf-\underline{\textbf{P}}aced \textbf{L}earning for Software \underline{\textbf{V}}ulnerability \underline{\textbf{D}}etection). 
\splvd dynamically selects source code for model training based on the stage of training, which simulates the human learning process progressing from easy to hard. 
\splvd has a data selector that is specifically designed for the vulnerability detection task, which enables it to prioritize the learning of easy source code. 
Before each training epoch, \splvd uses the data selector to recalculate the difficulty of the source code, select new training source code, and update the data selector. 
When evaluating \splvd, we first use three benchmark datasets with over 239K source code in which 25K are vulnerable for standard evaluations.
Experimental results demonstrate that \splvd achieves the highest F1 of 89.2\%, 68.7\%, and 43.5\%, respectively, outperforming the state-of-the-art approaches. 
Then we collect projects from OpenHarmony, a new ecosystem that has not been learned by general LLMs, to evaluate \splvd further. 
\splvd achieves the highest precision of 90.9\%, demonstrating its practical effectiveness.\\
Replication package: \url{https://figshare.com/s/bef3211194fc18fe375e}. 
\end{abstract}


 
\maketitle

\section{Introduction}
Software vulnerabilities are a security issue in software systems that attackers can exploit to cause harm, including system crashes, privacy data leaks, or financial loss~\cite{chakraborty2022deep, kalouptsoglou2024vulnerability, steenhoek2023empirical}. For instance, the Stuxnet worm exploited multiple vulnerabilities to attack the industrial control systems of Iran's nuclear facilities, causing damage to physical equipment and resulting in significant economic and political impacts~\cite{Stuxnet}. According to statista~\cite{statista}, the number of newly discovered Common Vulnerabilities and Exposures (CVE~\cite{neuhaus2010security}) has continuously risen, and the public release of these vulnerabilities has led to the formation of a vast vulnerability database~\cite{cve}. To prevent security issues caused by software vulnerabilities, the vulnerability detection system plays a crucial role in software security~\cite{johnson2011guide}. 

Conducting static code analysis on the submitted code is one of the effective approaches for detecting software vulnerabilities~\cite{jang2014survey, tomas2019empirical}. These approaches can be broadly classified into three categories~\cite{kalouptsoglou2023software}: 
\begin{inparaenum}
    \item Rules-based approaches: These approaches are based on the vulnerability rules predefined by experts, which identify matching vulnerabilities by statically scanning the source code\footnote{Hereafter, source code refers to function-level source code.}~\cite{vassallo2020developers}. While they can directly locate specific lines of code, they are limited to known patterns, which limits their effectiveness in identifying new vulnerabilities~\cite{goseva2015capability, bhandari2021cvefixes}. 
    \item Machine/deep learning-based approaches. These approaches are based on the features extracted from the function-level source code~\cite{lin2020software, steenhoek2023empirical, dam2018automatic}, and convert them into vectors or graph representations and use different neural networks such as recurrent neural networks (RNNs) for classification models to automatically learn the relationship between vulnerability patterns and labels, eliminating the need for manual rules~\cite{li2018vuldeepecker, li2021sysevr, chakraborty2022deep, zhou2019devign}. However, these approaches have low accuracy rates and can only provide rough predictions at the file or function level~\cite{seas2024automated, zheng2021vu1spg, cheng2019static}. It is difficult to determine the specific location of the vulnerability, which results in the lack of usability of these approaches in production~\cite{liu2019deepbalance, lin2020deep, steenhoek2024dataflow}.
    \item Fine-tuning pre-trained model-based approaches. Implementing vulnerability detection based on the pre-trained model obtained through training on large-scale source code, such as CodeBERT~\cite{feng2020codebert} and UniXcoder~\cite{guo2022unixcoder}, has become popular. These approaches adapt them to vulnerability detection tasks through transfer learning~\cite{zhou2024large}. They are capable of capturing code semantics and contextual dependencies effectively, supporting predictions such as line-level detection, which provides significant assistance for practical applications~\cite{fu2022linevul, kalouptsoglou2025transfer, zhang2023vulnerability, thapa2022transformer}. However, they are reliant on the code patterns learned from pretrained source code, and the quality of the training data influences their performance~\cite{cheng2021deepwukong, hanif2022vulberta, hin2022linevd}.
\end{inparaenum}

Despite the various approaches for vulnerability detection that exist, they place greater emphasis on how to extract the characteristics of the source code and construct the model~\cite{qiu2024vulnerability, cao2022mvd}, but neglect the selection of training data. Some research has shown that the vulnerability datasets are generally of low quality~\cite{ding2024vulnerability, liu2022investigating}. Many vulnerability datasets suffer from issues such as inconsistent labels, duplicate data, and erroneously classified source code~\cite{croft2023data}, which can lead to model bias in fine-grained predictions~\cite{li2010negative, jain2023code}. The current approaches lack a training approach to filter out high-quality training data. The models are affected by noisy data, which limits their accuracy in practice~\cite{steenhoek2024closing, chakraborty2024revisiting}. Therefore, they need an approach that can dynamically select training data to help the model prioritize the learning of high-quality data.

To this end, we propose a new approach: \textbf{\splvd} (\underline{\textbf{S}}elf-\underline{\textbf{P}}aced \textbf{L}earning for Software \underline{\textbf{V}}ulnerability \underline{\textbf{D}}etection). The core idea of \splvd is a data selector based on \spl, which enables the detection model to learn the source code from easy to hard during the training process, thereby reducing the influence of low-quality source code on the model's learning in the early stage. The key of the data selector lies in defining the difficulty of the source code and dynamically selecting the training code according to the training state. Specifically, the difficulty of each source code is calculated based on the model's confidence in its prediction and the correctness of the prediction. The data selector then uses a difficulty threshold to select the training code, which is automatically updated based on various features according to the training state. Before each training epoch, \splvd uses the data selector to recalculate the difficulty of each source code. Based on the updated difficulty, the selector then selects new training source code and updates the difficulty threshold accordingly, allowing the training data to be dynamically selected along with the training process of the model.


We evaluate \splvd on four datasets (BigVul, Devign, ReVeal, and OpenHarmony). \splvd outperforms all baselines on F1: on BigVul \splvd achieves an F1 of 89.2\%, together with 98.8\% accuracy and 96.1\% precision; on Devign it obtains accuracy of 74.8\% and an F1 of 68.7\%; and on the more challenging dataset \ie ReVeal, it still reaches an F1 of 43.5\%. \splvd achieves the highest F1 in the most common CVE categories. The ablation study indicates that incorporating \spl improves model performance: the F1 on BigVul, Devign, and ReVeal increased by 2.4\%, 3.6\%, and 2.0\%, respectively. Finally, \splvd achieves a precision of 90.9\% on OpenHarmony, demonstrating its effectiveness in real-world applications. 

The main contributions of this article are as follows.
\begin{itemize}
    \item \textbf{Approach}: A dynamic \spl approach based on training state feedback.
    \item \textbf{Model}: A vulnerability detection model that is trained using \spl. 
\end{itemize}

The remainder of this article is organized as follows. \Cref{SEC:MOT} describes the motivating example. \Cref{SEC:APP} presents the proposed \splvd. \Cref{SEC:EXP} and \Cref{SEC:RES} elaborate on experimental designs and results on evaluating \splvd, respectively. \Cref{SEC:DIS} discusses implications and limitations. \Cref{SEC:REL} reviews related work. Finally, we present threats to validity in \Cref{SEC:TTV} and conclude this article in \Cref{SEC:CON}.

\section{Motivating Example and Background}
\label{SEC:MOT}
This section first presents two examples in the real-world vulnerability dataset to motivate our work, and then explains why \spl is effective for vulnerability detection tasks even when the dataset contains noisy or erroneous source code.

\subsection{Motivating Example}
We found two types of data in the vulnerability dataset that might be wrongly labeled: (1) unrelated source code, and (2) non-modified source code. Then, we combined relevant studies to emphasize the significance of approaches for automatically filtering out low-quality data.

\begin{figure}[!htbp]
    \centering
    \includegraphics[width=\linewidth]{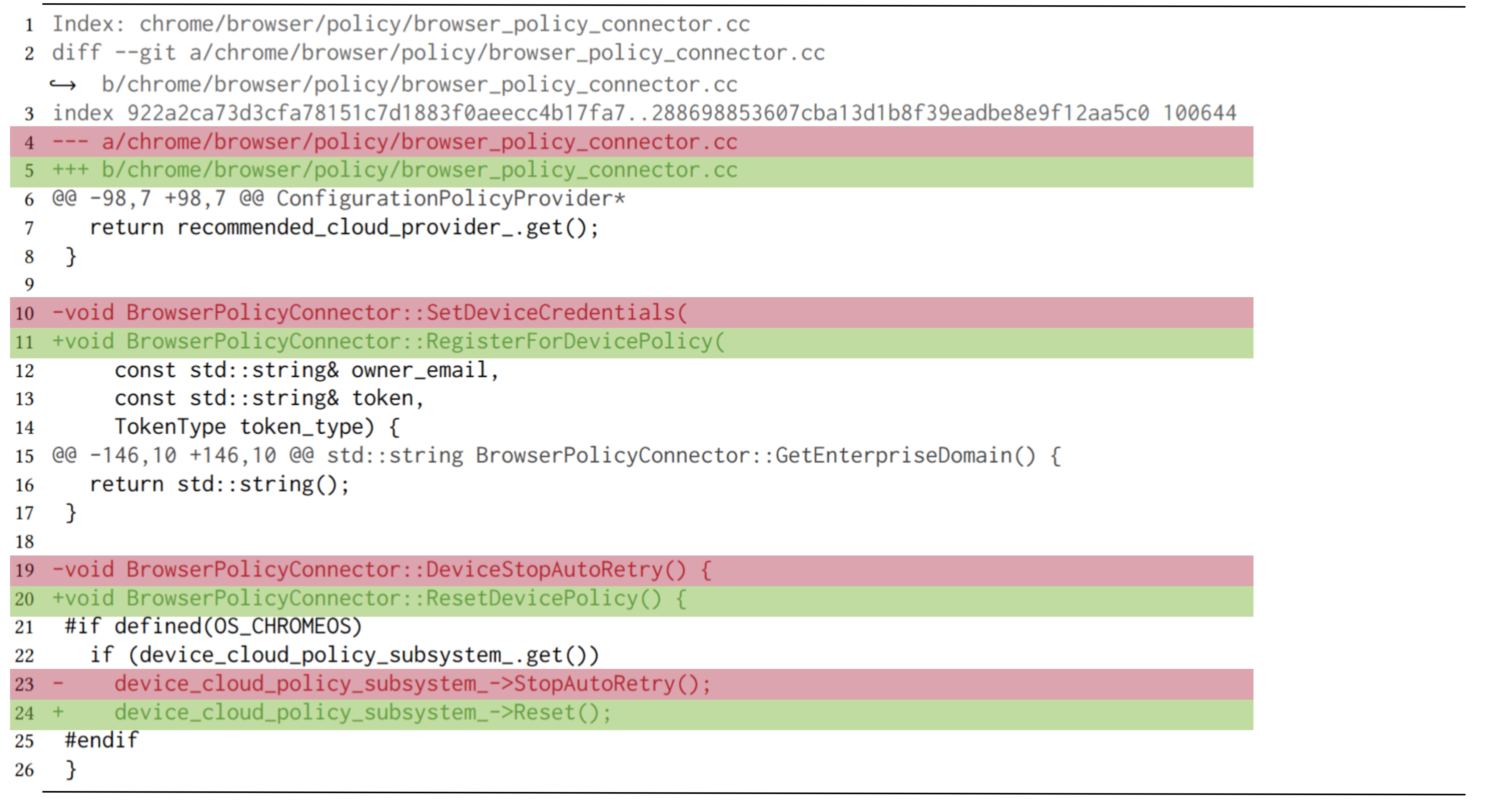}
    \caption{The Unrelated Code Is Wrongly Labeled}
    \label{FIG: example1}
\end{figure}

As illustrated in \Cref{FIG: example1}, an example in the BigVul dataset shows a case where unrelated code is wrongly labeled. In this example, only some method names have been modified (\eg \textit{SetDeviceCredentials} has been changed to \textit{RegisterForDevicePolicy}). Based on the current code context, it is difficult to determine if there are any vulnerabilities. However, in the dataset, it is labeled as vulnerable source code, and the type of the vulnerability is marked as CWE-399 (Improper Management of System Resources), which clearly has no relation to the code modification. Based on our analysis, the reason for this problem lies in the way the dataset is collected, which is by tracking the reference links of CVE. However, due to incorrect reference links or the fact that multiple areas are modified in a single commit, not every area is related to the vulnerability.

\begin{figure}[!htbp]
    \centering
    \includegraphics[width=\linewidth]{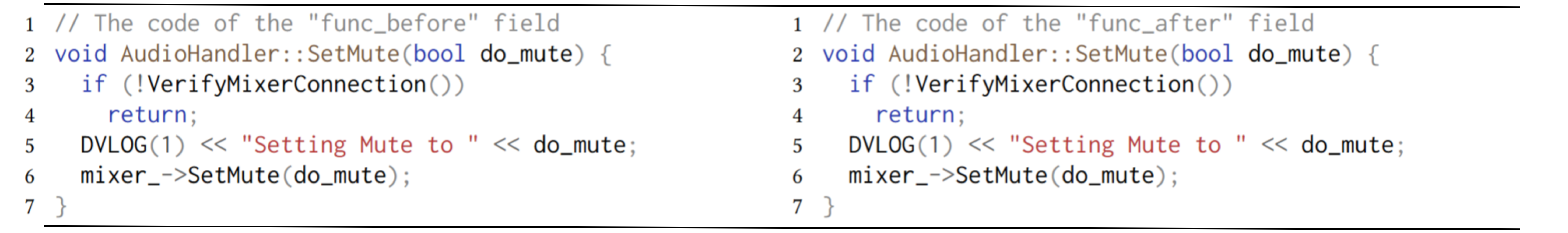}
    \caption{The Non-Modified Code Is Marked as Vulnerable}
    \label{FIG: example2}
\end{figure}

Then, we illustrate the potential problems of the dataset from the perspective of the collection approach of the BigVul~\cite{fan2020ac} dataset. The BigVul dataset is obtained by crawling the code modifications related to vulnerabilities in real-world project commits, thereby extracting the code that may contain vulnerabilities. However, we discovered nine data items in the dataset where the code remained unchanged throughout, but was wrongly labeled as vulnerabilities. The example shown in \Cref{FIG: example2} is one of the code labeled as having vulnerabilities, but without any code modifications. This function only performed the connection check and forwarded it to the \textit{mixer}. It does not parse the external input or perform any dangerous memory operations. Therefore, this code is more like an error extraction rather than a vulnerability. The reason for this problem might be that the dataset failed to accurately track the commits during the collection process.

In addition to the problems we identify, relevant studies have shown that the datasets in the field of vulnerability detection are generally of poor quality. ~\citet{croft2023data} pointed out that many current vulnerability datasets suffer from issues such as inconsistent labels, duplicate data, and erroneously classified source code. \citet{nie2023understanding} indicated that when noise is introduced into the vulnerability dataset, the effectiveness of various vulnerability detection approaches has significantly declined. These indicate that the quality of the dataset has a considerable impact on the detection performance of the model, especially in the early stage of training. The model is forced to learn from a full dataset that contains a great amount of noise, resulting in slow convergence and potential overfitting to noisy features, which leads to a high error rate in real-world applications. Therefore, in this work, we aim to design an adaptive data selection approach based on \spl. The core motivation is to quantify source code difficulty and dynamically adjust the selection of the training data, allowing the model to select high-quality data autonomously. This approach enables the model to first master core patterns from ``clean'' data and then gradually introduce more hard source code, thereby reducing noise interference and improving the model's learning efficiency and stability.

\subsection{Technical Preliminaries}
\Spl is a machine learning strategy that mimics the human autonomous learning process~\cite{bengio2009curriculum}, where students adjust their learning pace and curriculum based on their mastery of knowledge. Its design enables the model (``student'') to autonomously select training samples (``curriculums'') from easy to hard (first learning easy and reliable samples, and then gradually introducing hard and challenging samples), thereby improving model performance and generalization capability~\cite{kumar2010self}. It is highly adaptable to vulnerability detection tasks because the quality of the vulnerability dataset is poor, and \spl can reduce the impact of noise samples~\cite{tullis2011effectiveness}.

There are two key parameters in \spl: \emph{difficulty} and \emph{age}~\cite{wang2021survey}. The difficulty is determined by using the sample training loss of the current model as the standard for measuring the difficulty of the samples. The age is a parameter used to control the learning pace and determines the proportion of the easiest selected samples at each training epoch. Therefore, leveraging \spl requires designing the measure and update approaches for these two key parameters based on the specific task. The difficulty measurer determines the relative difficulty of samples, providing reference information for sample selection. The age updater updates the age parameter based on the training state and adjusts the proportion of training samples selected.

\begin{figure}[!htbp]
    \centering
    \includegraphics[width=\linewidth]{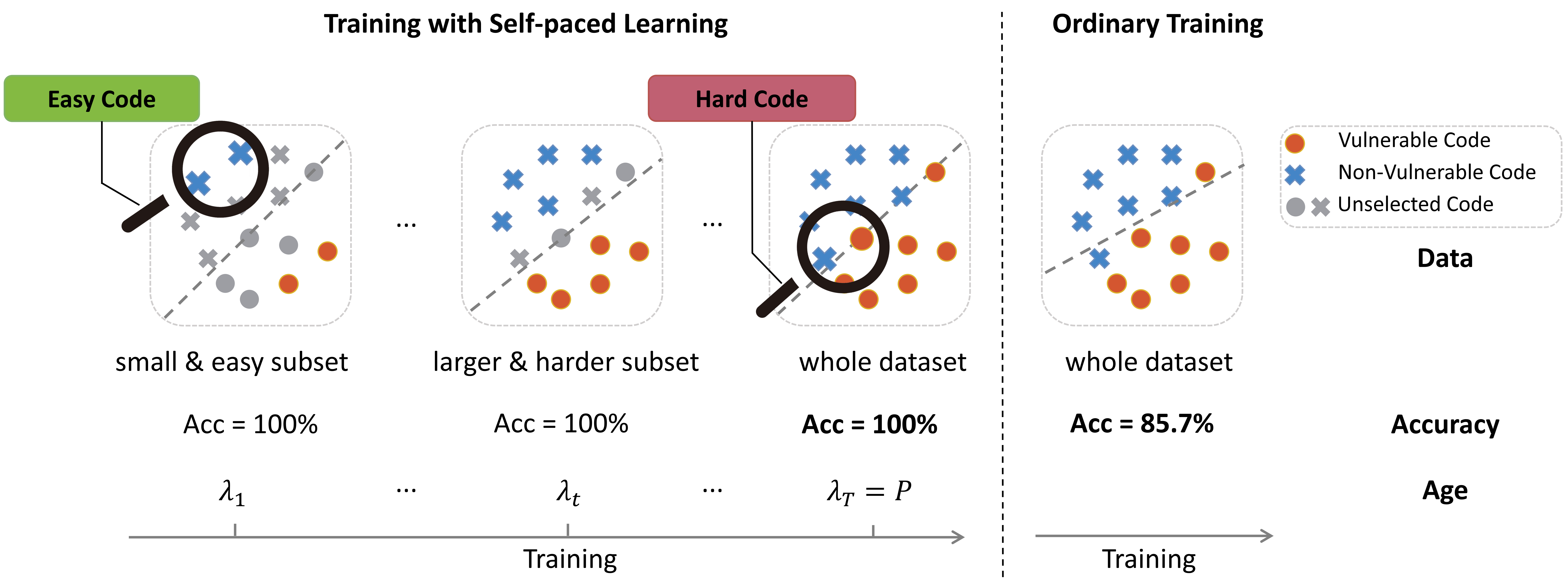}
    \caption{An Illustration of Self-Paced Learning for Model Training} 
    \label{FIG: SPL}
\end{figure}

The workflow of \spl is illustrated in \Cref{FIG: SPL}. Common training directly learns from the entire training set, which contains a large amount of hard source code and noise. This causes the model to fall into local optima at the early stages, resulting in relatively low accuracy (85.7\%). In contrast, \spl mimics the human learning process by starting with a ``small and simple subset'', allowing the model to quickly grasp easily distinguishable patterns in the early phase. As training progresses, the model gradually introduces harder subsets, thereby continuously enhancing its capability to handle hard cases. By the time the model is capable of processing the entire dataset, it is still able to maintain high accuracy (100\%).


\section{Approach: \splvd}
\label{SEC:APP}
This section first outlines \splvd, then presents the backbone of the model, the selection of training data, and the \spl training algorithm, respectively.

\subsection{Overview}
\begin{figure}[b!]
    \centering
    \includegraphics[width=0.99\linewidth]{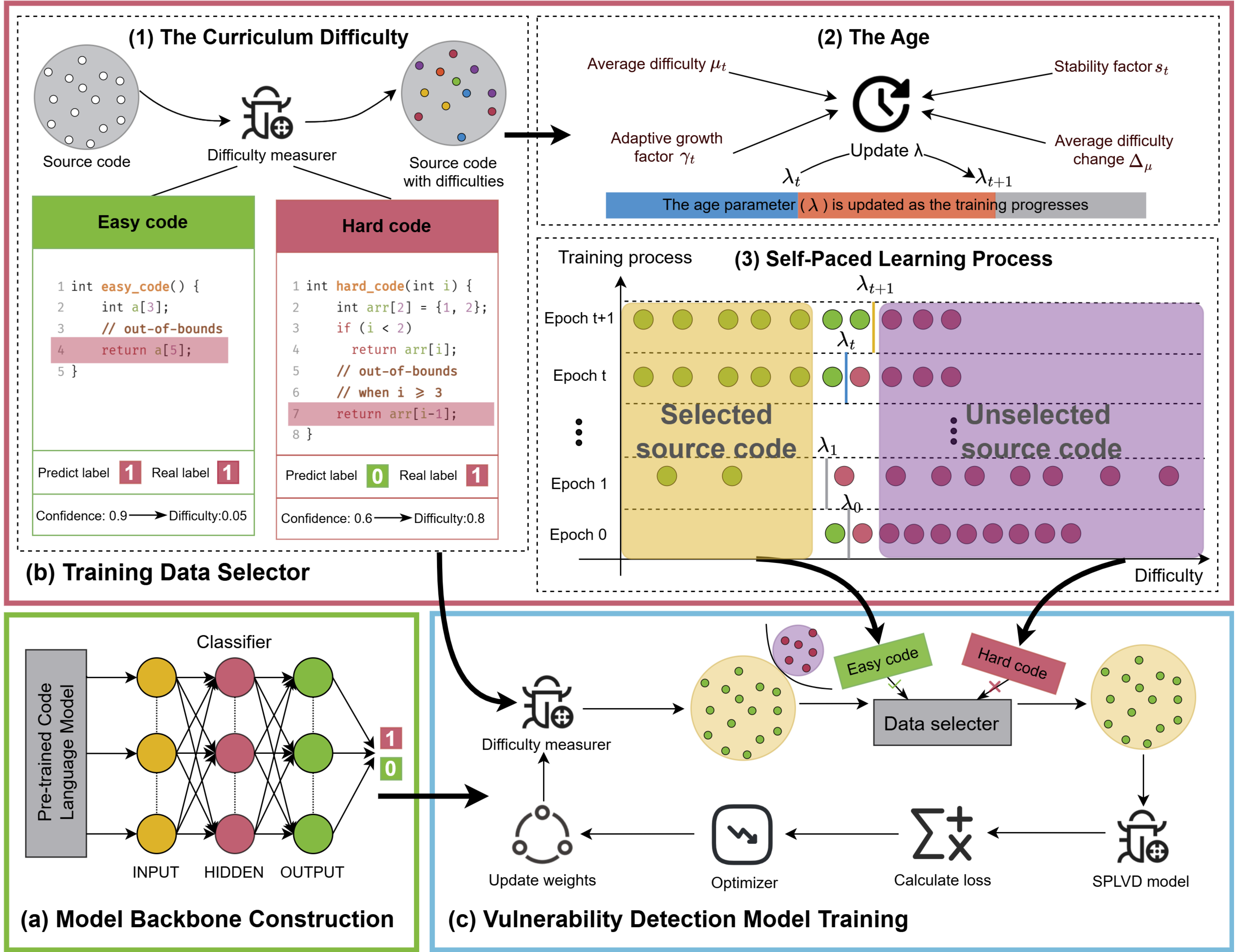}
    \caption{An Overview of the Proposed SPLVD}
    \label{FIG:main_approach}
\end{figure}

The overview of \splvd is shown in ~\Cref{FIG:main_approach}. Built upon a pre-trained code language model, \splvd incorporates \spl to select training source code.
\splvd first calculates the difficulty of all source code and initializes the age parameter. Then \splvd uses the data selector to select source code with lower difficulty as the training data.
Unlike traditional approaches that perform fixed-batch learning, \spl operates within the data selector during training. As shown in~\Cref{FIG:main_approach} (b), \spl dynamically chooses the source code whose difficulty is lower than the current age parameter, ensuring that at each stage the model learns from data most suitable for its current learning capability.
During each training epoch, the difficulty of the source code is recalculated, and the age parameter is updated. Then, the selected training data is used for model parameter weight updates. This process is repeated continuously until the \spl process is completed.
Overall, \splvd consists of three parts:
(1) the model backbone constructed for vulnerability detection,
(2) the training data selector based on \spl, and
(3) a training algorithm tailored for \spl.

\subsection{Model Backbone Construction}
The model backbone consists of two components: (1) the pre-trained code language model, and (2) the classifier layer for vulnerability detection.

\textbf{Pre-trained Code Language Model.}
We use UniXcoder~\cite{guo2022unixcoder} as the pre-trained code model for feature extraction. UniXcoder integrates code comments and AST, enabling the detection of syntax errors (e.g., null pointers, array out-of-bounds) and logical vulnerabilities by comparing code with comments. Its one-to-one AST mapping preserves structural information better than traditional models, reducing missed detections of hidden source code vulnerabilities.

\textbf{Classifier.}
We use RNN as the classifier layer~\cite{li2018vuldeepecker} for its compatibility with the context-sensitive nature of vulnerability detection: (1) its gating mechanism and sequence modeling capture long-range code dependencies that static classifiers such as fully connected layers cannot effectively learn; (2) building on sequence-level of UniXcoder syntactic and semantic representations, it further models local and global dynamic dependencies, augmenting the detection of context-sensitive vulnerabilities and improving classification accuracy.

\textbf{Model Backbone Construction.}
We build \splvd using UniXcoder as the base model and LSTM as the classifier. First, \splvd utilizes the tokenizer of Unixcoder to convert the source code into token ID sequences. Sequences longer than 512 tokens are truncated and shorter ones zero-padded for consistent input length. The token IDs are processed by Transformer encoder of the UniXcoder~\cite{vaswani2017attention}, and the ``[CLS]'' mark of token token-pooled output provides a global feature vector representing the overall semantics of the source code. The feature vector is then expanded into a sequence and fed into a single-layer RNN, which captures long-range dependencies. Finally, a fully connected layer compresses the RNN output into a logits vector, and a sigmoid function produces the vulnerability probability for binary cross-entropy loss optimization.

\subsection{Training Data Selector}
\label{SEC:selector}
To mitigate the impact of noisy or low-quality examples on model training, \splvd adopts \spl that gradually selects training source code ``from easy to hard'', enabling the model to learn more robustly. \Spl process is guided by two key factors: \emph{(1) the curriculum difficulty of training data} and \emph{(2) the age of the vulnerability detection model}. Both the source code difficulty and the age are updated by \textit{the self-paced update algorithm} throughout training, and their relative values determine whether source code is included in each epoch.

\textbf{The Curriculum Difficulty.}
For each source code $x_i$ with a true label $y_i \in \{0, 1\}$ (where 1 indicates a vulnerability and 0 indicates no vulnerability), let $p_{i1} \in [0, 1]$ represent the probability that the model predicts $x_i$ as vulnerability source code, and $p_{i0} \in [0, 1]$ represent the probability that the model predicts $x_i$ as non-vulnerability source code (both are outputs from the model's final sigmoid layer). \splvd defines the confidence difference $conf_i$ as the difference between the two class prediction probabilities:
$
conf_i = \text{abs}(p_{i0} - p_{i1})
$
where $conf_i \in [-1, 1]$, and the larger the absolute value, the stronger the model's confidence in the prediction (vulnerability or non-vulnerability). The source code difficulty $d_i$ is calculated based on $conf_i$ and prediction correctness.
\begin{equation}
d_i = 
\begin{cases} 
\frac{1 - conf_i}{2} & \text{if } \hat{y}_i = y_i \quad (\text{correct prediction}) \\
\frac{1 + conf_i}{2} & \text{if } \hat{y}_i \neq y_i \quad (\text{incorrect prediction})
\end{cases}
\end{equation}

where $\hat{y}_i = \text{argmax}(p_{i1}, p_{i0})$ is the predicted label of models for $x_i$. This difficulty definition ensures that $d_i \in [0, 1]$, with smaller values indicating lower difficulty. For correctly predicted source code, the difficulty decreases as the prediction confidence $conf_i$ increases (\eg for source code with $conf_i = 0.9$ and a correct prediction, $d_i = 0.05$, making it easy source code); for misclassified source code, the difficulty increases as confidence grows (\eg for source code with $conf_i = 0.9$ but a wrong prediction, $d_i = 0.95$, due to the model's ``overconfident error'', making it high-difficulty source code). This design mimics the learning pattern of humans. Source code that is correctly identified with high confidence is considered ``easy''. At the same time, those who mislead the model and make it overconfident are hard source code. As the model trains on the source code, its capability to fit the source code improves, and the difficulty of the source code gradually decreases.

\textbf{The Age.}
The decision of selection of source code depends on its relationship with the age of the vulnerability detection model (denoted as $\lambda_t$ in the $t$-th epoch). The age parameter controls the difficulty threshold for training, dynamically adjusting throughout the training process. The age ensures that the model gradually fits more hard source code as it improves. For initialization, $\lambda_0$ is set based on the difficulty distribution of all training source code. Specifically, it is initialized to the difficulty value at a certain quantile of the sorted training source code difficulties, defined by $r_{init}$. This ensures that the model starts training with the easy source code.

\textbf{Self-Paced Learning Process.}
The curriculum difficulty and the age parameter $\lambda$ are updated by \spl during training the model backbone in \splvd. At each epoch $t$, $\lambda$ is updated as follows.

\begin{equation}
\lambda_{t+1} = \lambda_t + \gamma_t \cdot (1 - \mu_t) \cdot s_t + \Delta_\mu
\end{equation}
where $\gamma_t$ is the adaptive growth factor, $\mu_t$ is the average difficulty, $s_t$ is the stability factor, $\Delta_\mu$ is the average difficulty change: $\Delta_\mu = \mu_t - \mu_{t-1}$. The definitions of $\gamma_t$ and $s_t$ are as follows.

\begin{equation}
\gamma_t = \gamma_0 \cdot \left(1 + \alpha \cdot (1 - r)\right)\quad
s_t = \frac{1}{1 + k \cdot \sigma_t}
\end{equation}
where $\gamma_0$ is the initial hyperparameter of the growth factor, $\alpha$ is a hyperparameter that controls the growth rate of $\gamma$, $r$ is the proportion of the currently selected source code to the total number of source code, $\sigma_t$ is the standard deviation of source code difficulty, and $k$ is the stability coefficient hyperparameter.

These parameters respectively have the following functions. The adaptive growth factor $\gamma_t$ controls the update rate of $\lambda$. When fewer source code are selected ($r$ is small), $\gamma_t$ increases, encouraging the inclusion of harder source code. When $r$ is high, the growth slows to prevent early usage to hard source code. The stability factor $s_t$ adjusts the update speed according to the stability of the distribution of difficulties. When $\sigma_t$ is smaller, $s_t$ becomes larger, thereby enabling $\lambda$ to grow at a faster rate. The average difficulty change. This allows the update to adapt to changes in overall training difficulty. 
The inclusion of $\Delta_\mu$ at the end serves to ensure that the update of $\lambda$ can adapt to the significant changes in overall training difficulty throughout the entire training process, preventing the changes from causing the \spl to fail.

In addition, to avoid the problem of unstable training causing too much hard source code to be added within one epoch, we limit the number of newly added source code in each epoch. This restriction ensures that the proportion of selected source code does not grow too quickly, helping maintain training stability and reduce overfitting. Specifically, during each training step, we calculate the proportion of the currently selected source code and based on this, determine how much new source code to add in the next step. However, there is an upper limit to the number of additions, and it cannot exceed a pre-defined growth rate $r_{max}$. In this way, new and more challenging source code will be gradually introduced into the training process, without experiencing significant changes.

Our dynamic age ($\lambda$) update algorithm is summarized in \Cref{algo:age-update}. By integrating the adaptive growth factor $\gamma$, stability factor $s$, and difficulty change $\Delta_\mu$,  the model progressively includes more source code ``easy-to-hard'' while maintaining stable training dynamics. The source code selection constraint further reinforces this process by controlling the increase in speed of age.

\begin{algorithm}[htbp]
\scriptsize
\caption{Age Parameter Update in Self-Paced Learning}
\label{algo:age-update}

\SetKwInOut{Input}{Input}\SetKwInOut{Output}{Output}
\SetKwProg{Fn}{Function}{}{}
\SetKwComment{tcc}{\(\rhd\)}{}

\Input{
Current age parameter $\lambda_t$; current selected ratio $r_t$; \\
Mean difficulties $\mu_t$, $\mu_{t-1}$; difficulty set $D_t$; \\
Growth factor $\gamma_0$; adjustment factor $\alpha$; \\
Max increment ratio $r_{\max}$; stability coefficient $k$
}
\Output{Updated age parameter $\lambda_{t+1}$}

\Fn{\texttt{updateLambda}($\lambda_t, r_t, \mu_t, \mu_{t-1}, D_t$)}{

$\gamma_t \leftarrow \gamma_0 \cdot \left(1 + \alpha (1 - r_t)\right)$ \tcc*{Adaptive growth factor}  

$\Delta_\mu \leftarrow \mu_t - \mu_{t-1}$ \tcc*{Difficulty change}  

$\sigma_t \leftarrow \text{std}(\text{subset near } \lambda_t)$ \tcc*{Local std near threshold}  

$s_t \leftarrow \frac{1}{1 + k \cdot \sigma_t}$ \tcc*{Stability factor}  

$\lambda' \leftarrow \lambda_t + \gamma_t (1 - r_t) \cdot s_t + \Delta_\mu$ \tcc*{Proposed update}  

$r' \leftarrow \frac{|\{d_i \in D_t \mid d_i \le \lambda'\}|}{|D_t|}$ \tcc*{New selected ratio}  

\If{$r' - r_t > r_{\max}$}{
  $r_{\text{target}} \leftarrow r_t + r_{\max}$ \;
  $\lambda' \leftarrow \text{quantile}(D_t,\ r_{\text{target}})$ \tcc*{Re-adjust via quantile}
}

$\lambda_{t+1} \leftarrow \min(1,\ \max(0,\ \lambda'))$ \tcc*{Clip to $[0, 1]$}

\Return $\lambda_{t+1}$
}
\textbf{EndFunction}
\end{algorithm}

\subsection{Vulnerability Detection Model Training}
The main components of training algorithms are:
(1) the definition of the loss function, and 
(2) the selection of optimization approaches.

\textbf{Loss Function.}
The optimization objectives of \spl include the model parameters $\theta$ and the self-paced weights (data selection variable) of the samples $v=[v_1,\dots,v_N]^\top\in[0,1]^N$. Dividing the training set into $M$ batches, then let the average loss of the $j$-th batch be $L_j(\theta) = \frac{1}{B_j}\sum_{i\in\text{batch }j}\ell_i(\theta)$ (where $\ell_i$ is the single-sample cross-entropy). Introduce a batch selection variable $v_j \in \{0,1\}$ and a age parameter $\lambda$~\cite{wang2021survey}. The joint training objective of \splvd is as follows.

\begin{equation}
\min_{\theta,\;v\in\{0,1\}^M}\; \mathcal{E}(\{d\}\mid \theta,\lambda)
\;=\; \sum_{j=1}^M v_j\,L_j(\theta)\;-\;\lambda\sum_{j=1}^M v_j
\label{formula:loss}
\end{equation}
where $\{d\}$ represents the difficulty list of each batch calculated using the difficulty calculation algorithm in \Cref{SEC:selector}. Notably, $v_j$ depends on the sample difficulty and the current age parameter:
$v_j = 1$ if and only if $d_j < \lambda$, otherwise $v_j = 0$.
That is, \splvd selects ``easy'' batches (with difficulties below the threshold $\lambda$) in the early training stage, gradually introduces more hard batches as $\lambda$ increases, thereby achieving sample scheduling from easy to hard and reducing the impact of sample noise on training.

\textbf{Optimizer.}
To solve \Cref{formula:loss}, we use AdamW~\cite{loshchilov2018decoupled} as the optimization algorithm for \spl, where the model starts by training on easy source code to quickly converge and gradually learn more hard ones. \Spl requires high optimization stability and strong generalization ability. AdamW, an extension of Adam, decouples weight decay from the gradient update, preventing instability caused by the coupling of L2 regularization~\cite{hoerl1970ridge} and adaptive learning rates in traditional Adam. This calculation allows AdamW to maintain rapid convergence while reducing overfitting, thus supporting both training stability and performance in \spl. The main formula of the AdamW algorithm is as follows.

\begin{equation}
\theta_t=\theta_{t-1}-\eta{\left(\frac{\hat{m}_t}{\sqrt{\hat{v}_t}+\varepsilon}+\lambda_{\mathrm{wd}}\theta_{t-1}\right)}\:,\quad\hat{m}_t=\frac{m_t}{1-\beta_1^t},\quad\:\hat{v}_t=\frac{v_t}{1-\beta_2^t}
\end{equation}
where $m_t=\beta_1 m_{t-1}+(1-\beta_1)g_t$ and $v_t=\beta_2 v_{t-1}+(1-\beta_2)g_t^2$, $\eta$ is the base learning rate, $\varepsilon$ is a small constant to prevent division by zero, and $\lambda_{\mathrm{wd}}$ is the weight decay coefficient. Using AdamW algorithm can combine adaptive step size (adjusted by $\hat v_t$) with momentum direction (provided by $\hat m_t$) and apply $L_2$-type regularization. The optimizer achieves a balance between convergence stability and generalization capability during the easy-to-hard training process required by \spl.

\section{Experimental Design}
\label{SEC:EXP}
This section presents the designs in our experiment, including research questions, datasets, baselines, evaluation metrics, and hyperparameter settings.

\subsection{Research Questions}
We propose five research questions (RQs) to guide the experimental evaluation of \splvd.
\smallskip

\begin{itemize}
   \item [\textbf{RQ1:}] How effective is \splvd in terms of overall performance? (Overall Performance)
   \item [\textbf{RQ2:}] How effective is \splvd in terms of CVE categories? (Detailed Analysis)
   \item [\textbf{RQ3:}] How important is self-paced learning of \splvd? (Ablation Study)
   \item [\textbf{RQ4:}] How generable is self-paced learning of \splvd? (Generalization Capability)
   \item [\textbf{RQ5:}] How effective is \splvd in real-world application? (Case Study)
\end{itemize}

\subsection{Datasets}
We build the experimental datasets through three steps: (1) dataset selection, (2) processing, and (3) splitting. \Cref{tab: dataset} shows the basic information of each dataset.
\smallskip

\noindent\textbf{Step 1: Dataset Selection.}
We select three most commonly used datasets and devise a new one to investigate the usefulness of \splvd in practice.

\begin{itemize}
    \item \textbf{BigVul} \cite{fan2020ac} consists of 180K source code from 348 open-source projects. Its key features include: (1) inclusion of both vulnerable and patched code; (2) coverage of 11 types of vulnerabilities in C/C++, \eg buffer and integer overflows; (3) provision of code change difference analysis. However, researchers have identified incorrect labels in this dataset.
    
    \item \textbf{Devign} \cite{zhou2019devign} consists of 27K source code from four open-source projects, such as FFmpeg and QEMU, and includes over, labelled through a cross-validation process based on security-related commits. This dataset utilizes Graph Neural Networks (GNNs) to capture program semantics and transforms source code into a Code Property Graph (CPG).

    \item \textbf{ReVeal} \cite{chakraborty2022deep} consists of 22K source code from two open-source projects. It is constructed through a multi-stage process: it uses the Joern tool to parse code into intermediate representations, employs graph neural networks to model code structures, and incorporates real-world vulnerability data (\eg CVEs) to enhance source code diversity. 

    \item \textbf{OpenHarmony}. To address RQ5, we develop a real-world dataset using OpenHarmony projects. We chose OpenHarmony because its code is relatively new and most of its content has not been utilized by pre-trained models, which can more realistically simulate the production environment. We selected several critical components (\eg arkcompiler\_ets\_runtime, arkui\_ace\_engine, and communication\_ipc) in OpenHarmony, which play essential roles in compilation, UI rendering, and inter-process communication. We then queried all reported CWE vulnerabilities from these projects, located the corresponding commit links via their references, and parsed the code change to extract potential vulnerable functions for annotation, thereby building the OpenHarmony dataset.
\end{itemize}

\begin{table}[htbp]
\captionsetup{font={bf}}
\footnotesize
\caption{Basic Information of the Experimental Datasets}
\begin{tabular}{lrrrrr}
\hline
\textbf{Dataset} & \textbf{\#Non-Vul.} & \textbf{\#Vul.} & \textbf{\#Projects} & \textbf{Published Year} & \textbf{Ref.}\\ \hline
BigVul & 177,736 & 10,900 & 348 & 2020 & \cite{fan2020ac} \\
Devign & 14,858 & 12,460 & 4 & 2019 & \cite{zhou2019devign} \\
ReVeal & 20,494 & 2,240 & 2 & 2022 & \cite{chakraborty2022deep} \\
OpenHarmony & 574 & 167 & 5 & 2025 & \cite{openharmony} \\ \hline
\end{tabular}
\label{tab: dataset}
\end{table}

\noindent\textbf{Step 2: Pre-Processing.}
We conducted data cleaning and processing on the inconsistent labels, duplicate data, and erroneously classified source code present in the dataset, reducing the influence of erroneous code on model training before training.

\noindent\textbf{Step 3: Splitting.}
Each dataset is divided into training, validation, and test sets in an 8:1:1 ratio. The split is fixed to ensure that the same data partitions are used for all approaches.

\subsection{Baselines}
We compare \splvd with eight state-of-the-art approaches that use pre-trained models. Note that VulGPT~\cite{kalouptsoglou2025transfer} has two implementations.
\smallskip

\begin{itemize}
    \item \textbf{CodeT5}~\cite{wang2021codet5} is a pre-trained model based on an encoder-decoder architecture that captures code characteristics and the correlation between natural language and code. It is fine-tuned for vulnerability detection, combining code comments to help identify logical vulnerabilities.

    \item \textbf{CodeBERT}~\cite{feng2020codebert} is a bimodal pre-trained model based on the Transformer architecture, capturing semantic relationships between natural language and code. After fine-tuning, it is effective for vulnerability classification and cross-modal code understanding.
    
    \item \textbf{UniXcoder}~\cite{guo2022unixcoder} is a unified cross-modal code model that integrates code, comments, and Abstract Syntax Tree (AST) information. It uses multi-task training to enhance robustness and capture dependencies related to vulnerabilities.
    
    \item \textbf{Starcoder2}~\cite{lozhkov2024starcoder} is an LLM optimized for long-context processing. It supports multi-objective training, making it highly effective in detecting vulnerability patterns in long code and across multiple programming languages. 

    \item \textbf{VulGPT}~\cite{kalouptsoglou2025transfer} is a vulnerability detection model based on pre-trained models. It utilizes full-parameter fine-tuning and word embedding extraction to demonstrate efficient transfer learning capabilities for vulnerability classification. 

    \item \textbf{EPVD}~\cite{zhang2023vulnerability} is a vulnerability detection model based on CodeBERT. It extracts code execution paths from the Control Flow Graph (CFG) and enhances feature fusion through CNN attention.

    \item \textbf{LineVul}~\cite{fu2022linevul} is vulnerability detection model based on Transformer architectures. It utilizes a two-stage process for function-level prediction and line-level localization without the need for external tools.

\end{itemize}

\subsection{Evaluation Metrics}
Following prior work~\cite{chakraborty2022deep, li2021sysevr, croft2023data}, we employ five metrics to evaluate each approach: Accuracy ($ACC$), Precision ($P$), Recall ($R$), F1 ($F1$), and Matthews Correlation Coefficient ($MCC$). $TP$ denotes the number of vulnerable source code that are detected as vulnerable, $FP$ denotes the number of source code that are not vulnerable but are detected as vulnerable, $TN$ denotes the number of source code that are non-vulnerable and are detected as not vulnerable, and $FN$ denotes the number of vulnerable source code that are detected as non-vulnerable.

\begin{itemize}
    \item \textbf{Accuracy} ($Acc = \frac{TP+TN}{TP+TN+FP+FN}$) measures the proportion of correctly classified source code among all source code.
    \item \textbf{Precision} ($P = \frac{TP}{TP+FP}$) measures the proportion of correctly predicted vulnerable source code among all source code predicted as vulnerable.
    \item \textbf{Recall} ($R = \frac{TP}{TP+FN}$) measures the proportion of correctly predicted vulnerable source code among all actual vulnerable source code.
    \item \textbf{F1} ($F1 = \frac{2 \times P \times R}{P + R}$) considers both accuracy and recall rate, reflecting the balance between detecting vulnerabilities and avoiding false positives.
    \item \textbf{Matthews Correlation Coefficient} ($MCC = \frac{TP\times TN - FP\times FN}{\sqrt{(TP+FP)(TP+FN)(TN+FP)(TN+FN)}}$) considers all four categories ($TP$, $TN$, $FP$, and $FN$), useful for imbalanced vulnerability datasets. $MCC$ values range between -1 and 1, with 1 being the optimal value.
\end{itemize}

We consider F1 to be the most critical evaluation standard. F1 is the most commonly used metric in vulnerability detection (\citet{kalouptsoglou2023software}), making it easier to compare with other research. The reason for choosing $MCC$ is that some datasets we selected are unbalanced, and it is one of the popular metrics used in related work~\cite{yao2020assessing}.

\subsection{Hyperparameter Settings}
For baselines such as LineVul, EPVD, and VulGPT, we adopt the best hyperparameters reported in the original paper or the replication packages. For \splvd, we perform a grid search to seek the optimal hyperparameters. The search space is defined following common practices and related work, including: (1) The proportion of initial selection codes ($r_{init}$): (0.05, 0.1, 0.15); (2) Stability Coefficient ($k$): (5, 10, 15); (3) The initial value of the growth factor ($\gamma_0$): (0.02, 0.025, 0.03); (4) The growth rate of $\gamma$ ($\alpha$) : (0.2, 0.3, 0.4); (5) The max value of $r$ growth ($r_{max}$):  (0.05, 0.1, 0.15). For each hyperparameter setting, we train the model and use F1 on the validation set to select the best hyperparameter setting. Training is terminated early if the validation loss does not improve for five consecutive epochs. \Cref{tab: Hyperparameter} presents the optimal hyperparameter.

\begin{table}[h]
\captionsetup{font={bf}}
\caption{Hyperparameter Settings for the Four Approaches}
\footnotesize
\begin{threeparttable}
\begin{tabular}{lrllrllrllr}
\hline
\multicolumn{2}{c}{\textbf{LineVul}} &  & \multicolumn{2}{c}{\textbf{EPVD}} &  & \multicolumn{2}{c}{\textbf{VulGPT}} &  & \multicolumn{2}{c}{\textbf{\splvd}} \\ \hline
epochs & 10 &  & epochs & 8 &  & max\_length & 512 &  & epochs & 50 \\
block\_size & 512 &  & block\_size & 400 &  & patience & 5 &  & max\_length & 512 \\
train\_batch\_size & 16 &  & train\_batch\_size & 40 &  & batch\_size\tnote{1} & 64 &  & batch\_size & 16 \\
eval\_batch\_size & 16 &  & eval\_batch\_size & 64 &  & batch\_size\tnote{2} & 8 &  & learning\_rate & 1e-5 \\
learning\_rate & 2e-5 &  & learning\_rate & 2e-5 &  & epochs\tnote{1} & 100 &  & $r_{init}$ & 0.1 \\
max\_grad\_norm & 1.0 &  & max\_grad\_norm & 1.0 &  & epochs\tnote{2} & 10 &  & $k$ & 10 \\
 &  &  & cnn\_size & 128 &  & learning\_rate \tnote{1} & 1e-3 &  & $\gamma_0$ & 0.025 \\
 &  &  & filter\_size & 3 &  & learning\_rate \tnote{2} & 2e-5 &  & $\alpha$ & 0.3 \\
 &  &  & d\_size & 128 &  &  &  &  & $r_{max}$ & 0.1 \\ \hline
\end{tabular}
\begin{tablenotes}   
    \footnotesize
    \item[1] It employs the word\_embedding approach.
    \item[2] It employs the fine-tuning approach.
\end{tablenotes} 
\end{threeparttable}
\label{tab: Hyperparameter}
\end{table}

We implemented \splvd using PyTorch. The experiments are performed on a machine with 1 NVIDIA GeForce RTX 4090 GPU and 1 vCPU Intel(R) Xeon(R) Gold 6459C.

\section{Result and Analysis}
\label{SEC:RES}
This section reports on results and analysis to answer research questions.

\subsection{RQ1: Overall Performance}
\textbf{Setup.}  
This experiment is evaluated on the standard test sets of three commonly used vulnerability detection datasets (BigVul, Devign, and ReVeal), comparing eight representative baselines with \splvd. Five evaluation metrics are adopted: Accuracy ($Acc$), Precision ($P$), Recall ($R$), Matthews Correlation Coefficient ($MCC$), and F1 ($F1$), with F1 serving as the primary comparison metric. We compare with eight baseline approaches, which can be broadly classified into three categories: using pre-trained models (CodeT5, CodeBERT, and UniXcoder), using Lora~\cite{hu2021lora} to fine-tune large language models (StarCoder2), and the state-of-the-art approaches from related works (VulGPT, EPVD, and LineVul). All approaches are executed under the same data splits and evaluation scripts, with the best metric for each dataset highlighted in bold for comparison (cf. \Cref{tab: RQ1}).

\textbf{Results.}
\splvd achieves the highest F1 across all three datasets, with particularly outstanding performance on BigVul, where it reaches 89.2\%, representing an improvement of about 2.9\% over the second-best approach VulGPT (fine-tuning) at 86.3\%. On BigVul, it achieves the highest Accuracy (98.8\%) and Precision (96.1\%), while maintaining a relatively high Recall (83.3\%), indicating that \splvd can effectively identify true vulnerabilities with very high precision. On Devign, \splvd achieves the best Accuracy (74.8\%) and F1 (68.7\%), which is about 1.7\% higher than VulGPT (67.0\%), and its overall Accuracy is 13.8\% higher, showing a clear advantage despite a relatively moderate Precision (56.1\%). On the more challenging ReVeal dataset, \splvd still achieves the highest F1 (43.5\%), about 1.4\% higher than VulGPT (42.1\%), with the improvement mainly attributed to its higher Recall (51.8\%, the best among all approaches). These results suggest that \splvd demonstrates stable improvements in F1 across different datasets, with particular strengths in enhancing recall and reducing false negatives.

For the $MCC$ metric, \splvd demonstrated the best performance across all datasets. On BigVul, its value reached 0.889, which was 0.03 higher than the second-best VulGPT (0.858). On both Devign and ReVeal, \splvd achieved the highest $MCC$ values of 0.338 and 0.368, respectively. These results indicate that, in addition to the F1 and Recall rate, \splvd achieves a more balanced classification between vulnerable source code and non-vulnerable source code. Another representative phenomenon is that certain models (\eg StarCoder2) achieved extremely high Recall (99.8\%) on Devign but at the cost of very low Precision, resulting in a very low $MCC$ value. This further highlights the necessity of balancing both Precision and Recall in vulnerability detection tasks.

\begin{table}[h]
\captionsetup{font={bf}}
\caption{The Overall Performance of Nine Approaches}
\scriptsize
\setlength{\tabcolsep}{1.15mm}
\begin{threeparttable}
\begin{tabular}{lccccccccccccccccc}
\hline
\multirow{2}{*}{\textbf{Approach}} & \multicolumn{5}{c}{\textbf{BigVul}} &  & \multicolumn{5}{c}{\textbf{Devign}} &  & \multicolumn{5}{c}{\textbf{ReVeal}} \\ \cline{2-6} \cline{8-12} \cline{14-18} 
 & $Acc$ & $P$ & $R$ & $MCC$ & $F1$ &  & $Acc$ & $P$ & $R$ & $MCC$ & $F1$ &  & $Acc$ & $P$ & $R$ & $MCC$ & $F1$ \\ \hline
CodeT5 & 0.980 & 0.889 & 0.756 & 0.810 & 0.817 &  & 0.649 & 0.632 & 0.552 & 0.288 & 0.590 &  & 0.902 & 0.504 & 0.286 & 0.331 & 0.365 \\
CodeBERT & 0.978 & 0.886 & 0.709 & 0.781 & 0.788 &  & 0.618 & 0.693 & 0.292 & 0.232 & 0.411 &  & \textcolor{blue}{\textbf{0.905}} & 0.569 & 0.147 & 0.255 & 0.234 \\
UniXcoder & 0.982 & 0.907 & 0.767 & 0.825 & 0.831 &  & 0.660 & \textcolor{blue}{\textbf{0.653}} & 0.543 & 0.309 & 0.593 &  & 0.904 & 0.553 & 0.116 & 0.222 & 0.192 \\
StarCoder2 & 0.953 & 0.889 & 0.035 & 0.171 & 0.067 &  & 0.457 & 0.456 & \textcolor{blue}{\textbf{0.998}} & -0.012 & 0.626 &  & 0.902 & \textcolor{blue}{\textbf{0.625}} & 0.022 & 0.105 & 0.043 \\
VulGPT\tnote{1} & 0.985 & 0.938 & 0.792 & 0.854 & 0.859 &  & 0.610 & 0.546 & 0.866 & 0.292 & 0.670 &  & 0.868 & 0.359 & 0.433 & 0.321 & 0.392 \\
VulGPT\tnote{2} & 0.985 & 0.937 & 0.800 & 0.858 & 0.863 &  & 0.652 & 0.624 & 0.600 & 0.297 & 0.611 &  & 0.900 & 0.488 & 0.371 & 0.366 & 0.421 \\
EPVD & 0.981 & 0.871 & 0.793 & 0.821 & 0.830 &  & 0.648 & 0.623 & 0.581 & 0.288 & 0.601 &  & 0.886 & 0.419 & 0.402 & 0.347 & 0.410 \\
LineVul & 0.984 & 0.892 & 0.828 & 0.851 & 0.859 &  & 0.644 & 0.600 & 0.659 & 0.289 & 0.628 &  & 0.896 & 0.467 & 0.380 & 0.365 & 0.419 \\ \hline
SPLVD & \textcolor{blue}{\textbf{0.988}} & \textcolor{blue}{\textbf{0.961}} & \textcolor{blue}{\textbf{0.833}} & \textcolor{blue}{\textbf{0.889}} & \textcolor{blue}{\textbf{0.892}} &  & \textcolor{blue}{\textbf{0.748}} & 0.561 & 0.887 & \textcolor{blue}{\textbf{0.338}} & \textcolor{blue}{\textbf{0.687}} &  & 0.867 & 0.374 & \textcolor{blue}{\textbf{0.518}} & \textcolor{blue}{\textbf{0.368}} & \textcolor{blue}{\textbf{0.435}} \\ \hline
\end{tabular}
\begin{tablenotes}   
    \footnotesize
    \item[1] It employs the word\_embedding approach.
    \item[2] It employs the fine-tuning approach.
\end{tablenotes} 
\end{threeparttable}

\label{tab: RQ1}
\end{table}

\begin{tcolorbox}[width=\linewidth, boxrule=0pt, sharp corners=all,
  left=2pt, right=2pt, top=2pt, bottom=2pt, colback=gray!20]
  \ding{45} \textbf{Answer to RQ1}: \splvd consistently achieves the best F1 across all datasets, demonstrating its effectiveness in balancing precision and recall for vulnerability detection. The highest $MCC$ value of \splvd indicates a better balance in identifying different types of source code.
\end{tcolorbox}

\subsection{RQ2: Detailed Analysis}
\textbf{Setup.}  
This experiment selects the top ten most frequent vulnerability categories in the BigVul dataset (classified by CWE-ID~\cite{CWE}) as the subjects of analysis (cf. \Cref{tab: RQ2}). For each type, we first pick out the vulnerable source code of this type from the BigVul dataset and then randomly select a certain proportion of non-vulnerable source code from all the non-vulnerable source code in the raw dataset so that the ratio of the selected vulnerable source code to the selected non-vulnerable source code is the same as that in the raw dataset. To ensure the reliability of the comparison, we exclude StarCoder2 since it performs extremely poorly in RQ1 on BigVul and yields an F1 of 0 on several CWE categories. The experiment compares \splvd with baselines in terms of F1 on each CWE category, to examine the applicability and performance gains of \spl.

\textbf{Results.}
\splvd outperforms baselines on most common CWE categories, with particularly notable advantages in several syntax/pattern-obvious vulnerability types, such as CWE-125 (Out-of-bounds Read), CWE-476 (NULL Pointer Dereference), CWE-190 (Integer Overflow or Wraparound), and CWE-189 (Numeric Errors). This demonstrates that \spl prioritizes ``easy-to-detect'' source code and gradually introduces harder ones to improve the model’s capability to capture vulnerability patterns with clear code characteristics. Meanwhile, in a few categories (\eg CWE-200 and CWE-264), baselines slightly outperform \splvd, suggesting that these vulnerabilities require stronger contextual reasoning or task-specific semantic knowledge, where a pre-trained model and fine-tuning strategy have an advantage.

For buffer overflow/out-of-bounds vulnerabilities (CWE-119/CWE-125), \splvd achieves a high F1 of 88.2\% and 88.3\%, respectively, showing an improvement of 1\%–3\% over most baselines. This indicates that when vulnerabilities exhibit clear local syntactic or data-flow features, \spl can better strengthen the model’s capability to detect such patterns. 
In contrast, for information disclosure/privacy-related CWE-200 and access control–related CWE-264, \splvd performs slightly worse than certain VulGPT variants. Upon examining the collected data, we found that a possible reason is that these categories rely on broader contextual information and semantics, where cross-function/cross-file semantics or external knowledge captured during pretraining play an important role.
In summary, \splvd demonstrates superior performance in most common vulnerability categories with distinct code fingerprints, but for vulnerability types that depend on deeper semantics or cross-resource information, further improvements may require incorporating stronger semantic pretraining or context-enhanced approaches.

\begin{table}[h]
\footnotesize
\captionsetup{font={bf}}
\caption{The Performance (F1) of Eight Approaches Regarding CVE Categories}
\scriptsize
\setlength{\tabcolsep}{1.5mm}
\begin{threeparttable}
\begin{tabular}{lcccccccccc}
\hline
\multirow{2}{*}{\textbf{Approach}} & \multicolumn{10}{c}{\textbf{CWE-ID}} \\ \cline{2-11} 
 & CWE-119 & CWE-20 & CWE-399 & CWE-125 & CWE-200 & CWE-264 & CWE-476 & CWE-190 & CWE-189 & CWE-362 \\ \hline
CodeT5 & 0.833 & 0.848 & 0.790 & 0.763 & 0.854 & 0.875 & 0.735 & 0.923 & 0.851 & 0.826 \\
CodeBERT & 0.808 & 0.775 & 0.756 & 0.714 & 0.784 & 0.800 & 0.640 & 0.815 & 0.735 & 0.708 \\
UniXcoder & 0.855 & 0.847 & 0.826 & 0.796 & 0.874 & 0.896 & 0.735 & 0.925 & 0.833 & 0.783 \\
VulGPT\tnote{1} & 0.863 & 0.852 & 0.848 & 0.828 & 0.893 & \textcolor{blue}{\textbf{0.917}} & 0.824 & 0.923 & 0.920 & 0.898 \\
VulGPT\tnote{2} & 0.874 & 0.867 & 0.844 & 0.863 & \textcolor{blue}{\textbf{0.925}} & \textcolor{blue}{\textbf{0.917}} & 0.824 & 0.913 & 0.875 & 0.875 \\
EPVD & 0.842 & 0.829 & 0.838 & 0.757 & 0.844 & 0.833 & 0.800 & 0.931 & 0.776 & 0.868 \\
LineVul & 0.857 & 0.856 & 0.833 & 0.862 & 0.871 & 0.878 & 0.755 & 0.912 & 0.846 & 0.846 \\ \hline
Our SPLVD & \textcolor{blue}{\textbf{0.882}} & \textcolor{blue}{\textbf{0.872}} & \textcolor{blue}{\textbf{0.852}} & \textcolor{blue}{\textbf{0.883}} & 0.881 & 0.891 & \textcolor{blue}{\textbf{0.830}} & \textcolor{blue}{\textbf{0.931}} & \textcolor{blue}{\textbf{0.923}} & \textcolor{blue}{\textbf{0.902}} \\ \hline
\end{tabular}
\begin{tablenotes}   
    \footnotesize            
    \item[1] It employs the word\_embedding approach.
    \item[2] It employs the fine-tuning approach.
\end{tablenotes} 
\end{threeparttable}
\label{tab: RQ2}
\end{table}

\begin{tcolorbox}[width=\linewidth, boxrule=0pt, sharp corners=all,
  left=2pt, right=2pt, top=2pt, bottom=2pt, colback=gray!20]
 \ding{45} \textbf{Answer to RQ2}: \splvd outperforms baselines on most common CWE categories, with particularly significant improvements in categories that exhibit clear syntactic or pattern-specific characteristics.
\end{tcolorbox}

\subsection{RQ3: Ablation Study}
\textbf{Setup.}  
To examine the contribution of \spl to model performance, this ablation study uses the baseline VD (UniXcoder as the pre-trained encoder and LSTM as the classifier) as the reference model, and constructs a variant \splvd by incorporating \spl into VD. The experiments are conducted on the BigVul, Devign, and ReVeal datasets, reporting the commonly used $\text{F1}$ as well as $\text{Top}_{250}\text{F1}$, $\text{Top}_{500}\text{F1}$, and $\text{Top}_{1000}\text{F1}$, which are computed by ranking source code based on model confidence and evaluating the top 250, 500, and 1000 predictions, respectively. The reason for choosing $\text{Top}_\text{N}\text{F1}$ as the metric is that in the actual application of the vulnerability detection model, the $\text{Top}_\text{N}$ data with higher prediction probabilities are regarded as potential candidates for vulnerabilities and are submitted to manual inspection. All models were trained and evaluated with the same data splits and evaluation scripts, and \Cref{tab: RQ3} presents both the values of each metric and the relative improvements (with the values in parentheses denoting the gains over VD).

\textbf{Results.}
As shown in \Cref{tab: RQ2}, introducing \spl into \splvd leads to overall improvements in F1 across all three datasets: on BigVul, F1 increased from 86.6\% to 89.1\% (+2.5\%); on Devign, from 65.1\% to 68.6\% (+3.5\%); and on ReVeal, from 41.5\% to 42.9\% (+1.4\%). For the confidence-ranked $\text{Top}_N$ metrics, BigVul and ReVeal demonstrated significant improvements in most $\text{Top}_N$ scores (\eg BigVul's $\text{Top}_{1000}\text{F1}$ rises from 95.9\% to 96.8\%; ReVeal improves by 3.7\%, 4.0\%, and 2.4\% on $\text{Top}_{250}$, $\text{Top}_{500}$, and $\text{Top}_{1000}$, respectively), indicating that \spl is effective in enhancing overall and high-confidence code detection effect.

On BigVul, \splvd improved the overall F1 by 2.5\% while increasing $\text{Top}_{1000}\text{F1}$ by 0.9\%, indicating that even when the high-confidence source code already has a good detection effect, \spl can still further optimize the balance between recall and accuracy on more data. On Devign, although slight decreases are observed in $\text{Top}_{250}$ and $\text{Top}_{500}$, \splvd still achieves a 3.5\% improvement in overall F1, suggesting that by incorporating more high-confidence (a high level of confidence implies a lower degree of difficulty.) source code, \spl effectively boosts detection capability on the full test set. On ReVeal, \splvd achieves consistent improvements across all $\text{Top}_N$ metrics (\eg a 4.0\% increase in $\text{Top}_{500}\text{F1}$), demonstrating that \spl brings more pronounced benefits in more challenging data domains. In summary, \spl dynamically selects training source code of varying difficulty, strengthening the model’s learning on easy cases, thereby improving overall F1 in most scenarios.

\begin{table}[h]
\captionsetup{font={bf}}
\caption{The Effectiveness (F1) of Self-Paced Learning on Three Datasets}
\scriptsize
\setlength{\tabcolsep}{2.0mm}
\begin{threeparttable}
\begin{tabular}{lcccc}
\hline
\multirow{2}{*}{Dataset} & \multicolumn{4}{c}{VD\tnote{1} - SPLVD\tnote{2}} \\ \cline{2-5} 
 & F1 & $\text{Top}_{250}\text{F1}$ & $\text{Top}_{500}\text{F1}$ & $\text{Top}_{1000}\text{F1}$ \\ \hline
BigVul & \textcolor{blue}{\textbf{0.866 - 0.891 (2.5\%$\uparrow$)}} & 1.000 - 1.000 (0) & \textcolor{blue}{\textbf{0.999 - 1.000 (0.1\%$\uparrow$)}} & \textcolor{blue}{\textbf{0.959 - 0.968 (0.9\%$\uparrow$)}} \\
Devign & \textcolor{blue}{\textbf{0.651 - 0.686 (3.5\%$\uparrow$)}} & \underline{0.984 - 0.978 (0.6\%$\downarrow$)} & \underline{0.910 - 0.900 (1.0\%$\downarrow$)} & \textcolor{blue}{\textbf{0.799 - 0.800 (0.1\%$\uparrow$)}} \\
ReVeal & \textcolor{blue}{\textbf{0.415 - 0.429 (1.4\%$\uparrow$)}} & \textcolor{blue}{\textbf{0.534 - 0.571 (3.7\%$\uparrow$)}} & \textcolor{blue}{\textbf{0.470 - 0.510 (4.0\%$\uparrow$)}} & \textcolor{blue}{\textbf{0.429 - 0.453 (2.4\%$\uparrow$)}} \\ \hline
\end{tabular}
\begin{tablenotes}   
    \footnotesize              
    \item[1] VD means using UniXcoder as the pre-trained model and using LSTM as the classifier.
    \item[2] SPLVD means training using \spl on the basis of VD.
\end{tablenotes} 
\end{threeparttable}
\label{tab: RQ3}
\end{table}

\begin{tcolorbox}[width=\linewidth, boxrule=0pt, sharp corners=all,
  left=2pt, right=2pt, top=2pt, bottom=2pt, colback=gray!20]
 \ding{45} \textbf{Answer to RQ3}: With the introduction of \spl, \splvd achieved improvements in overall F1 across all three datasets, and it shows an improvement in the identification of high-confidence source code.
\end{tcolorbox}

\subsection{RQ4: Generalization Capability}
\textbf{Setup.}  
To verify whether \spl is generally effective across different pre-trained models, we first select three pre-trained models (UniXcoder, CodeBERT, and CodeT5) based on the evaluation results from related works. We then conduct experiments on them, comparing two training strategies with and without \spl (SPL/NO-SPL). All experiments are conducted on the BigVul, Devign, and ReVeal datasets, reporting four evaluation metrics: Accuracy, Precision, Recall, and F1, with F1 serving as the primary comparison metric. The experimental results are presented in \Cref{tab: RQ4}.

\textbf{Results.}  
Using \spl consistently improved F1 across all three pre-trained models on all three datasets. For UniXcoder, using \spl raises F1 on BigVul, Devign, ReVeal from 86.6\% to 89.1\%, 65.1\% to 68.6\%, and 41.5\% to 42.9\%, respectively, with gains ranging from 1.4\% to 3.5\%. For CodeBERT, \spl improves BigVul and Devign by about 6.6\% and 28.9\%, respectively, showing that \spl can significantly enhance the capability of certain pre-trained models to identify source code. For CodeT5, \spl delivered substantial gains on BigVul (+16.0\%) and Devign (+13.1\%).

For CodeBERT, the most improvement appears on Devign, where F1 rises from 28.3 to 57.2 (+28.9\%), mainly driven by a dramatic increase in Recall (from 18.7\% to 68.6\%), indicating that \spl effectively helps the model capture more features. For CodeT5, F1 on BigVul jumped from 21.2\% to 37.2\% (+16.0\%), though accuracy dropped from 89.3\% to 83.5\%, suggesting that \spl significantly boosted recall of positive source code ($R$ increases from 25.0\% to 84.6\%), but at the cost of a decrease in accuracy, in exchange for a stronger detection capability.

\begin{table}[h]
\footnotesize
\captionsetup{font={bf}}
\caption{The Effectiveness of Self-Paced Learning Under Different Pre-Trained Models}
\scriptsize
\setlength{\tabcolsep}{0.9mm}
\begin{threeparttable}
\begin{tabular}{llccccccccccccccccc}
\hline
\multirow{2}{*}{\begin{tabular}[c]{@{}c@{}}\textbf{Pre-trained}\\ \textbf{Model}\end{tabular}} & \multirow{2}{*}{\textbf{Approach}} & \multicolumn{5}{c}{\textbf{BigVul}} &  & \multicolumn{5}{c}{\textbf{Devign}} &  & \multicolumn{5}{c}{\textbf{ReVeal}} \\ \cline{3-7} \cline{9-13} \cline{15-19} 
 &  & $Acc$ & $P$ & $R$ & $MCC$ & $F1$ &  & $Acc$ & $P$ & $R$ & $MCC$ & $F1$ &  & $Acc$ & $P$ & $R$ & $MCC$ & $F1$ \\ \hline
\multirow{2}{*}{UniXcoder} & SPL\tnote{1} & 0.988 & 0.947 & 0.841 & 0.887 & \textcolor{blue}{\textbf{0.891}} &  & 0.642 & 0.572 & 0.856 & 0.341 & \textcolor{blue}{\textbf{0.686}} &  & 0.875 & 0.389 & 0.478 & 0.362 & \textcolor{blue}{\textbf{0.429}} \\
 & NO-SPL\tnote{2} & 0.984 & 0.854 & 0.878 & 0.857 & 0.866 &  & 0.652 & 0.599 & 0.713 & 0.313 & 0.651 &  & 0.847 & 0.333 & 0.549 & 0.347 & 0.415 \\ \hline
\multirow{2}{*}{CodeBERT} & SPL & 0.892 & 0.231 & 0.372 & 0.238 & \textcolor{blue}{\textbf{0.285}} &  & 0.532 & 0.491 & 0.686 & 0.092 & \textcolor{blue}{\textbf{0.572}} &  & 0.730 & 0.159 & 0.406 & 0.118 & \textcolor{blue}{\textbf{0.228}} \\
 & NO-SPL & 0.810 & 0.144 & 0.462 & 0.176 & 0.219 &  & 0.568 & 0.583 & 0.187 & 0.105 & 0.283 &  & 0.406 & 0.122 & 0.813 & 0.109 & 0.212 \\ \hline
\multirow{2}{*}{CodeT5} & SPL & 0.835 & 0.238 & 0.846 & 0.393 & \textcolor{blue}{\textbf{0.372}} &  & 0.541 & 0.498 & 0.566 & 0.086 & \textcolor{blue}{\textbf{0.530}} &  & 0.706 & 0.176 & 0.540 & 0.172 & \textcolor{blue}{\textbf{0.266}} \\
 & NO-SPL & 0.893 & 0.185 & 0.250 & 0.158 & 0.212 &  & 0.581 & 0.576 & 0.306 & 0.136 & 0.399 &  & 0.654 & 0.162 & 0.603 & 0.162 & 0.255 \\ \hline
\end{tabular}
\begin{tablenotes}   
    \footnotesize       
    \item[1] SPL means using \spl based on pre-trained models.
    \item[2] NO-SPL means not using \spl in model training.
\end{tablenotes} 
\end{threeparttable}

\label{tab: RQ4}
\end{table}

\begin{tcolorbox}[width=\linewidth, boxrule=0pt, sharp corners=all,
  left=2pt, right=2pt, top=2pt, bottom=2pt, colback=gray!20]
  \ding{45} \textbf{Answer to RQ4}: \Spl enhanced F1 across all pre-trained models and datasets, with particularly notable gains for CodeBERT and CodeT5 on certain datasets. In most cases, the F1 improvement is primarily driven by increased recall.
\end{tcolorbox}

\subsection{RQ5: Case Study}
\textbf{Setup.}  
To validate the practicality of \splvd in real-world scenarios, we selected the open-source operating system \textit{OpenHarmony} as the target project. The experimental procedure was as follows: first, each trained model scanned the prepared source code and output its predicted vulnerable source code (denoted as \textit{Identified Count}). Next, these candidate vulnerabilities were reviewed by security experts, who confirmed the number of actual vulnerabilities (denoted as \textit{Confirmed Count}), from which precision was calculated. The information we have confirmed with the experts can be found at the replication package~\cite{replication_package}. It is worth emphasizing that this process strictly simulates the real industrial vulnerability detection workflow, which refers to automatic identification by the model, followed by expert verification and final confirmation. Among the models, StarCoder2 failed to detect any valid vulnerabilities on this dataset and is therefore reported as N/A.

\textbf{Results.}
As indicated in \Cref{tab: RQ5}, the performance of different approaches varies significantly in real-world projects. LineVul identified the largest number of vulnerabilities (25), but its confirmation rate was only 60\%, indicating that high detection came with a high false-positive rate. In contrast, \splvd identified fewer vulnerabilities (11), but 10 of them were confirmed, achieving a precision of 90.9\%, the best among all approaches. This result suggests that \splvd is more suitable as a ``high-confidence detector'' in practical engineering settings: although its overall number of findings may be lower than some coverage-oriented baselines, its outputs are more reliable, significantly reducing the manual verification workload of security experts. For comparison, UniXcoder achieved relatively high precision (85.7\%), but slightly lower than \splvd, showing that \spl provides additional benefits in reducing false positives and enhancing model stability. Overall, \splvd demonstrates superior practical value in real-world applications.

\begin{table}[h]
\captionsetup{font={bf}}
\caption{The Performance (Precision) of Eight Approaches on OpenHarmony}
\footnotesize
\begin{threeparttable}
\begin{tabular}{lccc}
\hline
\multirow{2}{*}{\textbf{Approach}} & \multicolumn{3}{c}{\textbf{OpenHarmony}} \\ \cline{2-4} 
 & \#Identified & \#Confirmed & Precision \\ \hline
CodeT5 & 17 & 13 & 0.765 \\
CodeBERT & 5 & 3 & 0.600 \\
UniXcoder & 14 & 12 & 0.857 \\
StarCoder2 & N/A & N/A & N/A \\
VulGPT\tnote{1} & 16 & 11 & 0.688 \\
VulGPT\tnote{2} & 16 & 13 & 0.813 \\
EPVD & 9 & 5 & 0.556 \\
LineVul & 25 & 15 & 0.600 \\ \hline
SPLVD & 11 & 10 & \textcolor{blue}{\textbf{0.909}} \\ \hline
\end{tabular}
\begin{tablenotes}   
    \footnotesize       
    \item[1] It employs the word\_embedding approach.
    \item[2] It employs the fine-tuning approach.
\end{tablenotes} 
\end{threeparttable}

\label{tab: RQ5}
\end{table}

\begin{tcolorbox}[width=\linewidth, boxrule=0pt, sharp corners=all,
  left=2pt, right=2pt, top=2pt, bottom=2pt, colback=gray!20]
  \ding{45} \textbf{Answer to RQ5}: \splvd achieved highest confirmation rate in the real-world validation. This indicates that it is more suitable as a high-confidence vulnerability detection tool, effectively reducing manual verification costs and enhancing reliability in practical applications.
\end{tcolorbox}

\section{Discussion}
\label{SEC:DIS}
This section first presents the implications we gained from the experiments, then explains how to use \spl in practice, and finally discusses some limitations of our approach.

\subsection{Implication}
The approach presents four key implications: (1) \splvd effectively measures low-quality source code as hard source code, proving the superiority of our difficulty measurer. (2) \Spl reduces missed detections of vulnerabilities by mitigating overfitting. (3) Model backbone influences absolute performance, with stronger pre-trained encoders yielding higher baselines and weaker ones gaining more from \spl. (4) Prediction threshold setting influences the precision/recall trade-off and should be appropriately selected based on business priorities.

\begin{figure}[!htbp]
    \centering
    \includegraphics[width=\linewidth]{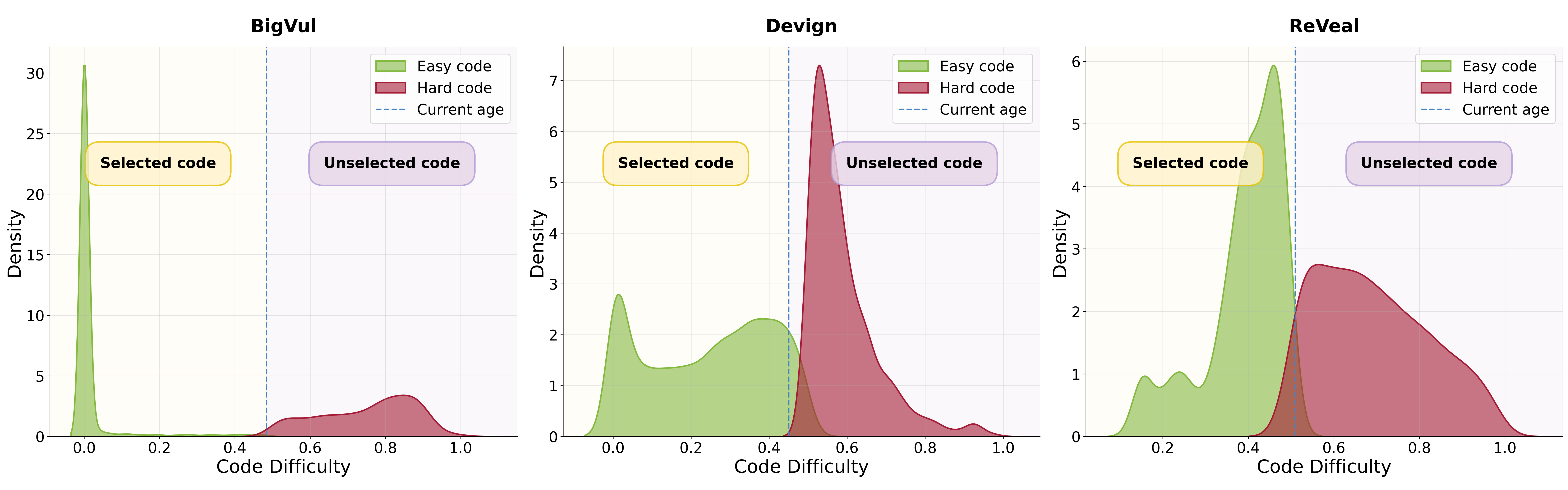}
    \caption{\splvd Measures the Distribution of Difficulty for Vulnerable Source Code on Three Datasets}
    \label{FIG:difficulty}
\end{figure}

\textbf{\splvd measures low-quality source code as hard.}
To investigate whether \splvd can effectively separate the easy and hard source code and give priority to learning the easy source code. We extract the difficulty calculation results of the last epoch on three datasets. Since the model learning mainly focuses on the vulnerable source code, we plot the difficulty distribution of the source code on the vulnerable source code as shown in \Cref{FIG:difficulty}. We can find that \splvd trained on the three datasets can effectively distinguish the easy source code from the hard source code, and the value of the final age parameter enables the model to select more easy source code for learning, thereby reducing the impact of hard source code on the model.

\begin{figure}[!htbp]
    \centering
    \includegraphics[width=\linewidth]{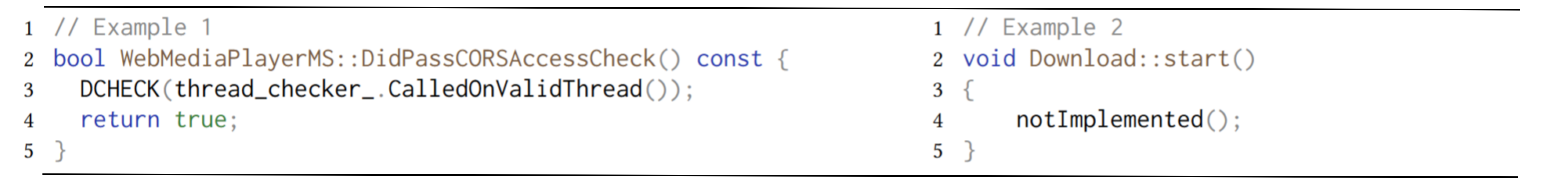}
    \caption{Two Examples of Short Contexts Found Among the Ten Hardest Source Code}
    \label{FIG: example3}
\end{figure}

In addition, we find that the source code that is ultimately labeled as hard has a high correlation with the low-quality data. This proves the superiority of our difficulty measurer. Taking the BigVul dataset as an example, all source code that we previously identified as potentially being mislabeled in \Cref{SEC:MOT} are found within the hard source code. We also select the ten hardest source code for manual analysis, and the result shows that no obvious vulnerabilities are found in these source code. At the same time, we find that among these source code, four of them are simple empty destructors wrongly marked as vulnerabilities. There are some examples, as shown in \Cref{FIG: example3}, which have very short contexts, and it is impossible to directly determine whether there are vulnerabilities or not. This once again proves that there is low-quality code in the vulnerability dataset, and our approach is capable of effectively filtering out such source code. The analysis results of the code can be found in the replication package~\cite{replication_package}.

\begin{figure}[htbp]
  \centering
  \includegraphics[width=\linewidth]{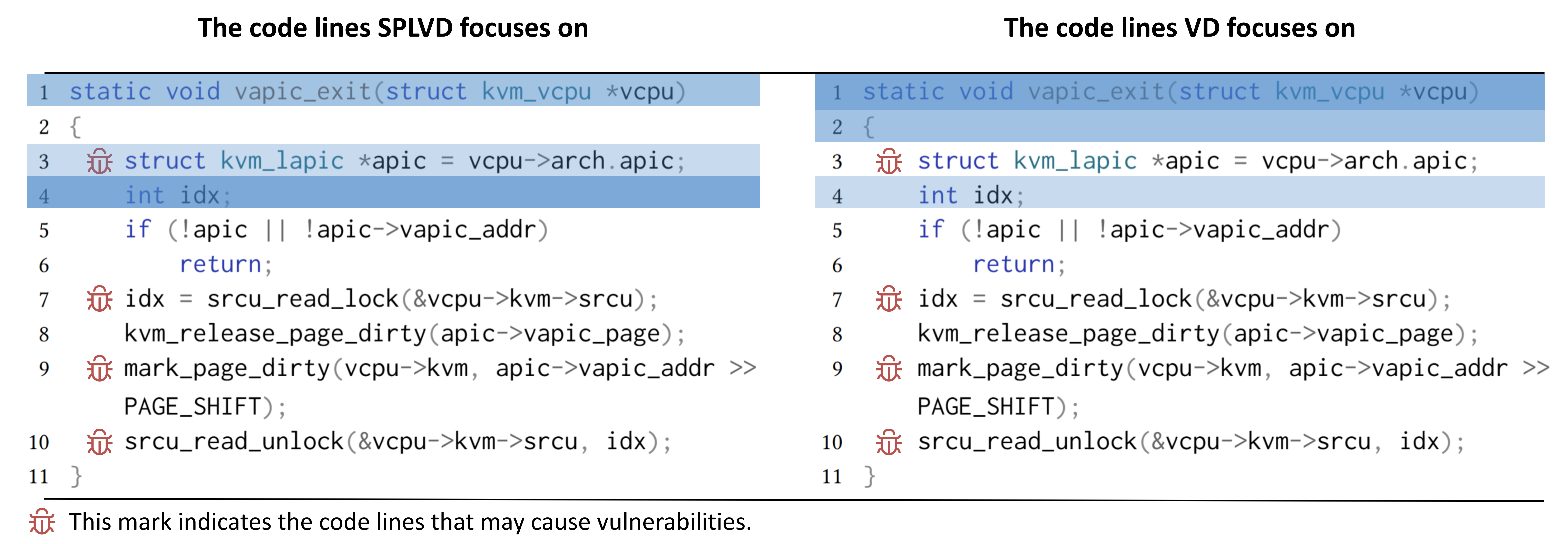}
  \caption{An Example Shows Models Trained with and Without Self-Paced Learning Focused on Different Rows}
  \label{FIG: example4}
\end{figure}

\textbf{Self-paced learning reduces missed detections of vulnerabilities.}
The experimental results of RQ1, RQ3, and RQ4 demonstrate that \spl consistently improves recall and F1 across different datasets and base models, confirming its effectiveness in enhancing vulnerability detection performance. These improvements can be attributed to two main factors: (1) the bespoke learning from easy to hard source code, and (2) dynamically introducing source code based on training age. 
By gradually learning from easy to hard source code, the model avoids early exposure to hard or noisy data, reducing overfitting and improving generalization, which helps it detect true vulnerabilities more effectively. 
Meanwhile, by dynamically introducing source code based on training age, \spl gradually selects diverse data into the training process, allowing the model to capture global patterns more comprehensively and preventing gradients from being dominated by a narrow subset of source code. We selected some examples from the BigVul dataset that were correctly labeled. By referring to the code lines that the model's attention focused on, we explain why the model trained using \spl has better detection performance than the original model. Take \Cref{FIG: example4} as an example; this example is labeled as CWE-20 (Improper Input Validation) in the dataset. The potential vulnerability lies in the fact that the function does not check whether ``vcpu'' is null before accessing it. We can observe that \splvd obtained through \spl can precisely focus its attention on the rows that may cause an exception, while the VD that is not trained with \spl fails to do and even focuses on the parentheses. This is the reason why its prediction results are incorrect.
Together, these mechanisms explain why \spl is particularly effective in boosting recall, thereby reducing the risk of missed vulnerabilities and improving system security.

\textbf{Model backbone influences detection performance.}
According to the results of RQ4, differences in the initial capability of pre-trained encoders affect both the final performance and the gain brought by SPL. With UniXcoder as the baseline, applying SPL already achieved stable and relatively high F1 (BigVul 89.1\%, Devign 68.6\%, ReVeal 43.5\%), with steady improvements. In contrast, for base models such as CodeBERT and CodeT5, SPL provided much larger relative gains (\eg CodeBERT on Devign improved F1 by 28.9\%, and CodeT5 on BigVul improved by 16\%). The results suggest that when the base model has weaker recall or representation capability, \spl can provide more significant improvements. In engineering practice, it is advisable to prioritize pre-trained models with stable performance and strong semantic representation to achieve higher absolute performance, while recognizing that \spl can serve as an effective enhancement strategy for weaker base models.

\begin{figure}[!htbp]
    \centering
    \includegraphics[width=\linewidth]{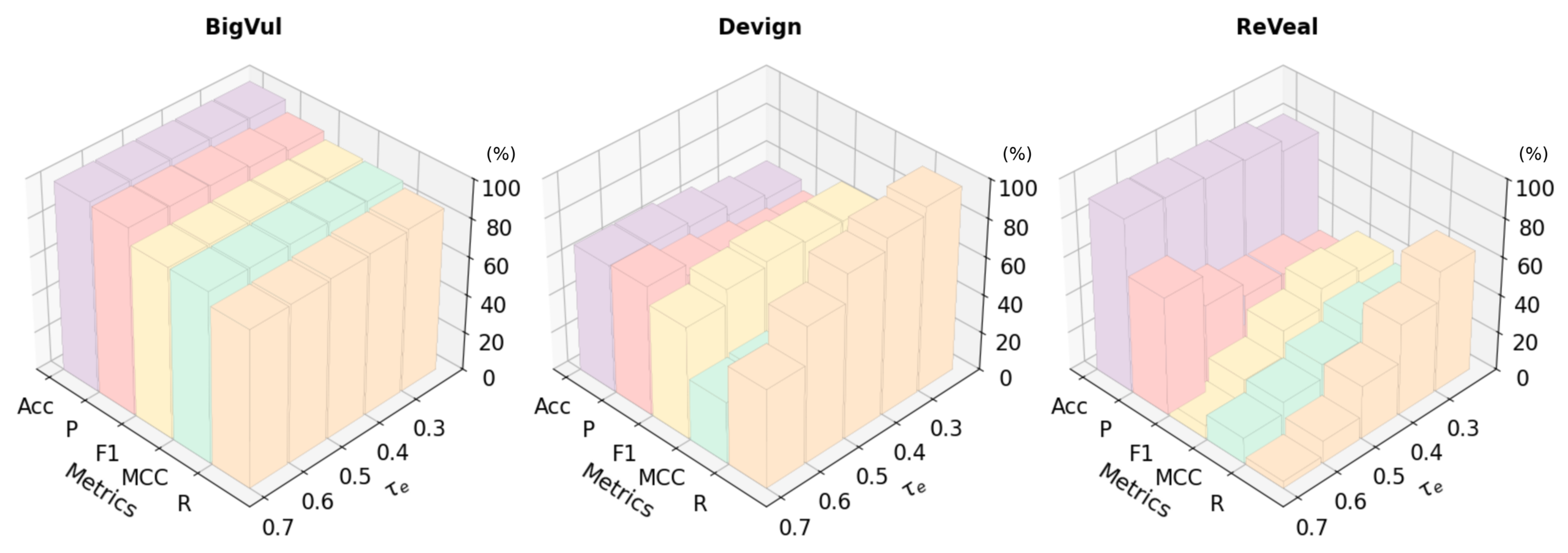}
    \caption{The Performance of \splvd Using Different Thresholds}
    \label{FIG:threshold}
\end{figure}

\textbf{Prediction threshold influences detection performance.}
Our experiments revealed that the choice of prediction threshold $\tau_e$ during testing has an impact on results. \Cref{FIG:threshold} shows the results in five evaluation metrics of \splvd under different thresholds on three datasets. It can be observed that, especially on the smaller-scale datasets of Devign and ReVeal, Precision ($P$) and Recall ($R$) rates decrease and increase, respectively, as the threshold value increases. The highest F1 corresponds to different threshold values on different datasets. It indicates that the selection of the threshold has an impact on the detection performance of the model. In practical applications, thresholds should be set according to business priorities: higher thresholds when prioritizing precision, and lower thresholds when prioritizing recall. Thresholds can be dynamically adjusted based on the project requirements. For high-risk source code, a ``low threshold + manual review'' workflow can be adopted to balance efficiency and reliability.

\subsection{Usage}
The approach can be used to train vulnerability detection models with \spl and then apply the trained model for automated vulnerability detection with expert review.

\textbf{Usage of Self-Paced Learning in Model Training.}
For researchers in the field of code vulnerability detection, \spl can be integrated into conventional model training workflows. For example, based on the UniXcoder pre-trained model and an LSTM classifier, source code difficulty is estimated from prediction confidence and correctness, and the selection threshold is dynamically adjusted using an age-parameter function. Only source code meeting the current difficulty standard is used for parameter updates, completing the \spl process. This process progresses automatically with training iterations, eliminating the need for manual intervention in source code selection and thereby significantly reducing dependence on high-quality annotated data.

\textbf{Usage of \splvd in Vulnerability Detection.}
For DevSecOps teams, vulnerability analysts, etc., the trained model can be directly applied to predict vulnerabilities in new source code. For source code predicted with high vulnerability probability, a human review process can be incorporated to leverage expert understanding of vulnerability context, compensating for the model's limitations in hard logical reasoning. This ``automated detection + manual verification'' paradigm not only takes advantage of \spl but also ensures the reliability of detection results.

\subsection{Limitations}
Though efforts have been made, \splvd still has two limitations in terms of \emph{training} and \emph{verification}.

\textbf{Longer Training Time.}
Due to the incorporation of additional source code difficulty estimation and dynamic training age updates in \spl, the overall training process is more complex. On large-scale datasets or with deep pre-trained models, the extended training time becomes more pronounced, which limits the applicability of \spl. Therefore, future work could explore optimizing the efficiency of difficulty calculation and source code selection to mitigate this issue while maintaining effectiveness.

\textbf{Require Manual Verification.}
Although \splvd achieved the highest F1 across multiple datasets, the model's precision remains insufficient in certain cases. This suggests that the model may still generate a non-negligible number of false positives. As a result, the current approach cannot fully replace manual inspection and needs to be complemented with human review. Future research should aim to further improve model precision across diverse datasets, driving vulnerability detection systems toward full automation and high reliability.

\section{Related Work}
\label{SEC:REL}
This section reviews related work from perspectives of training strategy, data selection, and approaches in vulnerability detection.

\subsection{Training Strategy}
Using appropriate training strategies plays a significant role in enhancing the model's performance. \spl has been proven to be beneficial in enhancing the performance and robustness of the model~\cite{yang2021mspld, zhang2023denoising}.
\citet{yuan2024self} incorporated \spl with dynamic difficulty measurement and scheduling into hierarchical multi-label classification, achieving superior performance on 20 datasets.
\citet{zhao2024symmetric} integrated \spl with a symmetric scheduler and gradient-based difficulty measurement into domain generalization, significantly improving cross-domain performance.
\citet{chen2024spcontrastnet} applied \spl with dynamic thresholding and staged sampling to few-shot text classification, constructing the Self-paced Contrastive Network, which outperformed state-of-the-art approaches.
\citet{chen2024self} used \spl with two-dimensional pseudo-label filtering and dynamic thresholding in semi-supervised text classification, constructing the Self-Paced Pairwise model, which outperformed the baseline approach and demonstrated excellent performance in alleviating overfitting and mislabeling problems.

\subsection{Data Selection}
Many studies have focused on the data quality problems existing in the commonly used vulnerability datasets~\cite{ding2024vulnerability, liu2022investigating}, which can lead to a decline in the performance of the model. For example,  
\citet{nie2023understanding} investigated the causes, impacts, and denoising effects of label errors on the synthetic and real datasets. The results show that label errors arise during the data collection and annotation stages, and 30\% of noise can cause an F1 drop of over 20\% for the model. 
\citet{croft2023data} adopted the standard five data quality attributes (accuracy, uniqueness, etc.) to conduct a quality assessment on four vulnerability datasets. They found that all datasets had quality problems. The accuracy rate of vulnerability labels in real-world datasets ranged from 20\% to 71\%, the duplication rate ranged from 17\% to 99\%, and these problems led to a 29\% to 80\% decrease in model precision.
\citet{dil2025towards} noticed that the vulnerability dataset has quality problems. By employing strategies such as LLM combined with generative knowledge prompts, they studied the impact of LLM filtering on data quality and the performance of vulnerability prediction models. After filtering, BigVul increased the accuracy rates of models by 7\% to 9\%. However, using LLM for data filtering is too costly, and its adaptability to different vulnerability detection models varies.

\subsection{Vulnerability Detection}
When building approaches for vulnerability detection, researchers have selected different source code features and machine learning models~\cite{zheng2020impact, kalouptsoglou2024vulnerability, purba2023software}. 
\citet{li2021sysevr} extracted four types of grammatical features from the abstract syntax tree, and integrated data flow and control flow semantics through the program dependency graph. They adopted the bidirectional gated recurrent unit as the model.
\citet{fu2022linevul} used a byte-pair encoding approach to segment sub-words, combined with the pre-trained CodeBERT model to generate word embeddings, and utilized the self-attention mechanism to achieve row-level positioning. 
\citet{zhang2023vulnerability} constructed a grammar control flow graph based on the abstract syntax tree and selected three representative execution paths using the greedy algorithm. They chose the pre-trained CodeBERT model to extract features and used a CNN to capture the correlation features. 
\citet{kalouptsoglou2025transfer} chose Word2Vec, FastText, and Bag-of-Words to extract code features, selected CodeBERT/CodeGPT2 to extract word embeddings or sentence embeddings. 
Recently, researchers have utilized LLMs for vulnerability detection.
\citet{lu2024grace} used Joern to extract code attributes, employed CodeT5 to extract semantic features, and combined them with highly relevant code examples. 
\citet{yang2025dlap} extracted the multi-form representations of the source code, the detection probabilities of the deep learning model, and the scan results of the static analysis tool. They then combined the two prompt techniques of Contextual Information Learning and Chain-of-Thought to achieve the adaptive optimization of LLM.

\section{Threats to Validity}
\label{SEC:TTV}
This section describes threats to validity and the efforts made to mitigate their effects.

\textbf{Internal Validity.}
We have made an effort to determine whether the performance gains of \splvd are attributable to \spl. To eliminate the difference from dataset splits, we predefine and reuse the same train/validation/test partition (ratio 8:1:1) across all experiments. We train and evaluate each baseline using its officially recommended hyperparameters (\eg learning rate, batch size) and use grid search to select the hyperparameters in \splvd to reduce the influence of parameter selection on vulnerability detection. When we explore the improvement effect of \spl, we also conducted experiments using multiple pre-trained models to prevent threats to validity caused by experiments on a single model.

\textbf{Construct Validity.}
We control construct validity via careful selection of datasets and evaluation metrics. 
For dataset selection, we use three widely adopted open-source datasets (BigVul, Devign, and ReVeal) based on the datasets chosen by several baseline approaches and further collect data from OpenHarmony to improve industrial representativeness. Some constructed datasets (\eg NVD~\cite{nvd}) are not selected because the purpose of \splvd is to improve the usability of vulnerability detection models; therefore, open-source datasets are more capable of testing the performance of \splvd.
For evaluation metrics, we select the five most common metrics in vulnerability detection. In the ablation study, to further compare the improvement in the detection performance of the model on high-confidence source code achieved through \spl, we introduce the evaluation metric $Top_NF1$. To maintain the completeness of VulGPT, we select the two best approaches proposed within it as baselines. In addition, in RQ2 we do not compare \splvd with Starcoder2. This is mainly because its detection results on BigVul are poor, with an F1 of 0 in many CWE categories, and thus lacks comparability.

\textbf{External Validity.}
To test the generalization capability of \splvd, our experiments are conducted on three well-known open-source datasets in vulnerability detection and a new one--OpenHarmony. When collecting OpenHarmony, we selected the key components repositories with a large scale (with 1K+ forks). We conducted experiments using various C/C++ datasets of different sizes (ranging from large to medium/small), with varying levels of class balance (balanced and unbalanced) and different project types. This diverse evaluation demonstrated the stability of \splvd under different data conditions and proved its strong generalization capability. We have disclosed the data and materials used to develop and evaluate \splvd to support its continuous improvement.

\section{Conclusion}
\label{SEC:CON}
In this article, we propose \splvd, a new approach to address the issue of low-quality data in software vulnerability detection. \splvd dynamically quantifies source code difficulty to achieve progressive learning and effectively reduces interference from noisy data. \splvd combines the pre-trained model UniXcoder with an LSTM classifier and designs a dynamic age parameter function based on training state feedback. Experimental results on three well-known datasets demonstrate that \splvd outperforms existing approaches. Moreover, we built a real-world dataset using source code from OpenHarmony to evaluate \splvd further and confirm the detection results with security experts. In summary, our work improves the practicality and reliability of vulnerability detection models through \spl. In the future, we plan to investigate more definitions of source code difficulty and validate \splvd in programming languages other than C++.

\bibliographystyle{ACM-Reference-Format}
\bibliography{splvd}

\end{document}